\documentclass[10pt,conference]{IEEEtran}
\usepackage{epsfig,setspace,amsmath,epsf,amssymb,bm,theorem,cite,graphicx, epstopdf,algorithm,algpseudocode,float,color,mathtools,authblk}
\usepackage[table,xcdraw]{xcolor}
\usepackage{booktabs}
\usepackage{subfigure}
\usepackage{bbm}
\usepackage{physics}
\usepackage[short,c2]{optidef}

\newtheorem{theorem}{Theorem}
\newtheorem{corollary}{Corollary}

\newtheorem{remark}{Remark}
\newtheorem{lemma}{Lemma}

\newcommand{\e}{{\mathbb{E}}}

\IEEEoverridecommandlockouts
\allowdisplaybreaks
\begin{document}

\title{Source Coding for a Wiener Process}

\author{Sahan Liyanaarachchi \qquad Ismail Cosandal \qquad Sennur Ulukus\\
	\normalsize Department of Electrical and Computer Engineering\\
	\normalsize University of Maryland, College Park, MD 20742 \\
	\normalsize \emph{sahanl@umd.edu} \qquad \emph{ismailc@umd.edu} \qquad \emph{ulukus@umd.edu}}

\maketitle

\begin{abstract}
    We develop a novel source coding strategy for sampling and monitoring of a Wiener process. For the encoding process, we employ a four level ``quantization'' scheme, which employs monotone function thresholds as opposed to fixed constant thresholds. Leveraging the hitting times of the Wiener process with these thresholds, we devise a sampling and encoding strategy which does not incur any quantization errors. We give analytical expressions for the mean squared error (MSE) and find the optimal source code lengths to minimize the MSE under this monotone function threshold scheme, subject to a sampling rate constraint. 
\end{abstract}

\section{Introduction}
Throughout the literature, Wiener processes or rather Brownian motion, has become a fundamental unit of study whose horizons expand from governing the statistical dynamics of particles to modeling the behavior of contemporary applications such as stock markets and computer networking \cite{brown_stocks,brown_networks}. Thus, the remote of estimation of a Wiener process has become a widely pursued research interest in recent past.

One of the first papers which explicitly studies the remote estimation of Wiener processes is \cite{Yin_Sun}, where the authors find the optimal sampling policy that should be employed when monitoring a Wiener process through a random delay channel, to minimize the mean squared error (MSE). In there, the authors show that a zero-wait (ZW) sampling policy is not optimal, and hence introduce a sample-and-wait policy which is proven to be optimal. Thereby, this study has spawned a multitude of its variants including its extensions to different processes such Ornstein-Ulhenbeck (OU) process, different channel statistics and multiple sources \cite{early_sample,Htang,Karim_Arafa}. All these works are interesting in their own right, however, they all share a fundamental drawback: They all involve the transmission of the real-valued samples of the process which cannot be practically implemented. Thus, the need for an appropriate quantization scheme arises naturally.

The work in \cite{Wei_chen} takes the need for quantization into account and provides analytical expressions for the MSE over a finite horizon. They resort to a high resolution quantization scheme and highlight the relations between the age process \cite{yates2020age}, quantization errors and the MSE. Their transmission strategy involves quantizing the difference of the Wiener process between two successive sampling points, where they show that this strategy leads to unbounded quantization errors as the time horizon is expanded. To circumvent this problem, they resort to a multi-level quantization error correcting scheme which involves quantizing the accumulated sum of quantization errors in multiple stages. However, this leads to another problem: In this procedure in order to transmit the $N^i$th sample, the transmitter needs to go through $i$ quantization stages before transmission. Thus, the computational complexity of this scheme becomes unbounded in the infinite time horizon.

In our work, we aim to completely eliminate the quantization errors by employing an event-driven sampling strategy. In particular, we take samples only when our process hits a known threshold. Similar to \cite{Wei_chen}, we too transmit the successive differences of the Wiener process between two sampling points. However, during the transmission of a sample, the Wiener process may very well be outside our predetermined thresholds. Hence, deviating from the norm, we employ two monotone function thresholds to effectively capture the process, if it goes outside the predetermined thresholds. This procedure is explained in detail in Section \ref{sec:mono}. We exploit the hitting times of the process with these thresholds and provide analytical expressions for the MSE, $\e[\Delta_{mse}]$, of the process defined by,
\begin{align}
    \e[\Delta_{mse}]=\limsup_{T\to \infty}{\frac{1}{T}\e\left[\int_{0}^T(W_t-\hat{W}_t)^2\,dt\right]},
\end{align}
where $W_t$ is the Wiener process and $\hat{W}_t$ is its estimate. Further, we develop a source coding scheme to minimize the  MSE under a given sampling rate. The introduction of source coding scheme makes the transmission delay dependent on the signal, adding a new layer of complexity to the problem in \cite{Yin_Sun}, where the transmission delays were independent of the process.

To summarize our contributions: (i) We introduce a novel event-driven sampling strategy with monotone function thresholds for monitoring a Wiener process. (ii) We provide analytical expressions for the MSE under this monotone function thresholding scheme. (iii) We find the optimal source codes for the event-driven sampling and encoding strategy.

\begin{figure}
    \centering
    \includegraphics[width=0.9\linewidth]{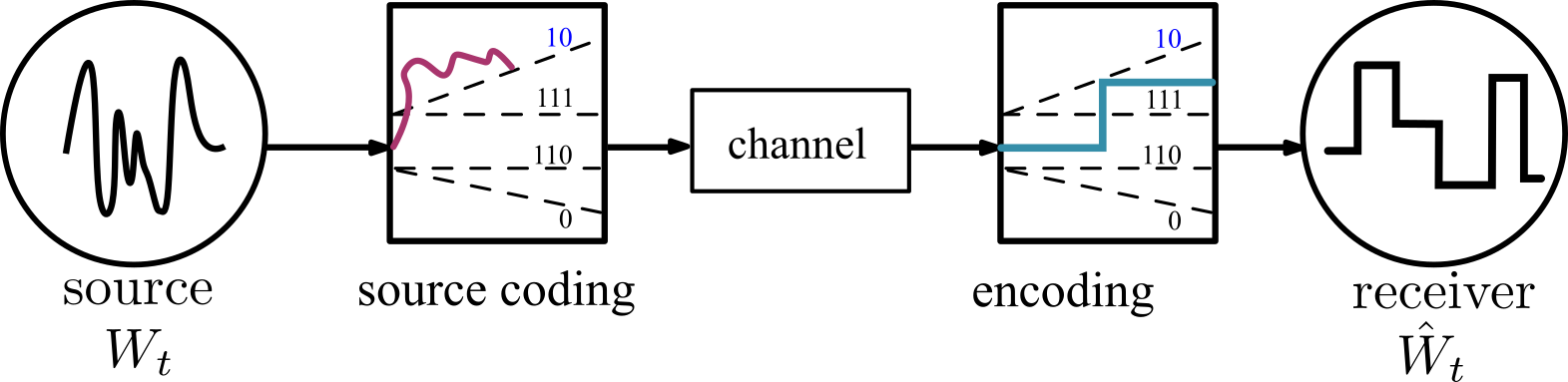}
    \caption{System model.}
    \label{fig:sys_model}
\end{figure}

\section{Related Work}
The work in \cite{Brownian_coding}  studies the problem of coding a Brownian motion in the interval of $(0,1)$ so as to minimize the supremum of the quantization error under a given norm. They achieve this by representing the Brownian motion through piecewise constant paths on intervals and encoding the exit times of the process from these intervals. However, since the exit times are also continuous, this too induces quantization errors. They overcome this by approximating the random variable associated with the exit times with an appropriate discrete random variable and treat its discrepancy as the quantization error. However, this study does not consider sampling or transmission processes into account.

The reference \cite{rate_distortion} looks into the problem of reconstructing a Wiener process from a quantized set of samples where they evaluate the trade-off between the sampling rate, bit rate and MSE. As opposed to per sample transmission, they sample the Wiener process multiple times and encode a vector of samples for transmission using a fixed number of bits each time.

In the realm of source coding, \cite{optimal_codes} analyzes the problem of  sampling and transmission of a discrete random source so as to minimize the age of information. They show that Shannon codewords constructed with respect to a tilted version of the probability mass function (PMF) of the random source is close to optimal (only constant away). This idea is further expanded by the work in \cite{selec_enc}, which selectively encodes a subset of the most probable realizations instead of encoding all possible realizations of the random source.

The closest to our work is \cite{tsi_wei_chen}, where the authors employ a threshold based sampling scheme along with timing side information (TSI) and a buffer. However, as opposed to our system, they flood the channel with a sample whenever the process exceeds the threshold even during an ongoing transmission. Additionally, they encode and transmit the inter-sampling times and simply use a single bit to denote which threshold the process hits. An issue in this approach is the delays incurred due to the backlog of packets. As highlighted in \cite{Yin_Sun}, sampling when the channel is busy is not optimal when monitoring a Wiener process. 

\section{Monotone Function Thresholds}\label{sec:mono}
Now, we describe our sampling and source coding scheme. Let $W_t$ denote a Wiener process  with unit variance. Assuming a bufferless system (no queue), a sampler will monitor this process and take samples when needed based on channel availability (i.e., sample only when channel is free). Once a sample is taken, it will be source coded and will be transmitted  to the remote monitor (see Fig.~\ref{fig:sys_model}). We assume that channel would incur a  delay of one unit of time to transmit one bit. Suppose we take the $n$th sample at time $S_n$ and it is coded into a length of $L_n$. Then, it would be delivered to the monitor at time $D_n=S_n+L_n$.

The monitor maintains an estimate of the Wiener process denoted by $\hat{W}_t$, which should be set to the value of the last received sample in order to minimize the MSE \cite{Yin_Sun}. At each sampling instance, we will not be encoding and transmitting the actual value of the Wiener process, but rather its deviation from the previous sampling instance. In other words, we will be transmitting successive increments of $W_t$ instead of the actual value of $W_t$.  In particular, for the $(n+1)$th sample, we will be encoding and transmitting  $Z_{n+1}=W_{S_{n+1}}-W_{S_n}$, instead of $W_{S_{n+1}}$.  Hence, at time $D_{n+1}$, the monitor can update its estimator to $\hat{W}_{D_{n+1}}=W_{S_{n+1}}=\sum_{i=1}^{n+1}Z_i$ assuming, $W_0=0$. The monitor will maintain this update until the next sample is received.

\begin{figure}
    \centering
    \includegraphics[width=0.9\columnwidth]{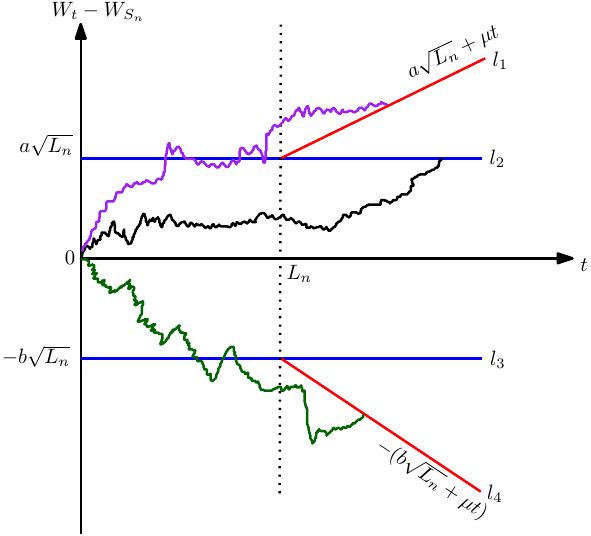}
    \caption{Monotone function thresholding scheme.}
    \label{fig:mono_quant}
\end{figure}

To encode the successive increments, we employ the sampling and encoding strategy illustrated in Fig.~\ref{fig:mono_quant}. In particular, to take the $(n+1)$th sample, we will observe the process $W_t-W_{S_n}$ and wait until it hits one of the thresholds depicted in Fig.~\ref{fig:mono_quant} after time $D_n=S_n+L_n$. In here $a$, $b$ are some non-negative constants. Note that since the Wiener process has independent increments, we have taken $S_n=0$ in Fig.~\ref{fig:mono_quant}. Let $X_n=W_{D_n}-W_{S_n}$. If $-b\sqrt{L_n}\leq X_n\leq a\sqrt{L_n}$, the next sample will be taken when $W_t-W_{S_n}$ hits either $a\sqrt{L_n}$ or $-b\sqrt{L_n}$. In this instance, we will be taking the next sample at the stopping time $\tilde{\tau}(X_n)=\min\{\tilde{\tau}_{a_n},\tilde{\tau}_{b_n}\}$, where $\tilde{\tau}_{a_n}=\inf\{t:W_{t+D_n}-W_{S_n}=a\sqrt{L_n}\}$ and $\tilde{\tau}_{b_n}=\inf\{t:W_{t+D_n}-W_{S_n}=-b\sqrt{L_n}\}$. Moreover, $\tilde{\tau}(X_n)$ is a well-behaved stopping time (i.e., $\e[\tilde{\tau}(X_n)]<\infty$). If the process hits $a\sqrt{L_n}$, we will be sending a codeword of $l_2$ bits and if it hits $-b\sqrt{L_n}$ will be sending a codeword of $l_3$ bits. In either case, we will be transmitting and encoding which event occurred as opposed to the actual value of the sample. However, upon decoding the codeword, the monitor can unveil the actual value of the sample to be either $a\sqrt{L_n}$ or $-b\sqrt{L_n}$.

If $X_n>a\sqrt{L_n}$, we will wait until $W_t-W_{S_n}$ hits the increasing threshold defined by $a\sqrt{L_n}+\mu t$. In this case, we will be taking a sample at the stopping time which is defined by $\hat{\tau_a}(X_n)=\inf\{t:W_{t+D_n}-W_{S_n}=a\sqrt{L_n}+\mu t\}$. This stopping time is equivalent in distribution to a hitting time of a standard Brownian motion having a drift of $\mu>0$, with a positive threshold. Therefore, this is also a well-behaved stopping time (see Corollary \ref{cor:mnts}). This event will be encoded by a separate codeword which will be transmitted to the monitor. Upon decoding this codeword, the receiver has the following pieces of information: At time $D_{n+1}$, it can deduce the exact sampling time of the $(n+1)$th sample as $S_{n+1}=D_{n+1}-L_{n+1}$ and thereby $\hat{\tau}_a(X_n)$ since it knows $D_n$. Next, from the received codeword, it knows that the sample was taken as result of process hitting the increasing threshold $a\sqrt{L_n}+\mu t$. Therefore, it can deduce the exact value of the successive increment using  $Z_{n+1}=a\sqrt{L_n}+\mu \hat{\tau}_a(X_n)$ without inducing any quantization errors. In this case, we will be using $l_1$ bits to encode this event. The case $X_n<-b\sqrt{L_n}$ follows a similar treatment where we define the hitting time of the increment process with the decreasing threshold $-(b\sqrt{L_n}+\mu t)$ as $\hat{\tau}_b(X_n)=\inf\{t:W_{t+D_n}-W_{S_n}=-(b\sqrt{L_n}+\mu t)\}$ and we will use $l_4$ bits to encode this event.

\begin{figure}
    \centering
    \includegraphics[width=0.9\linewidth]{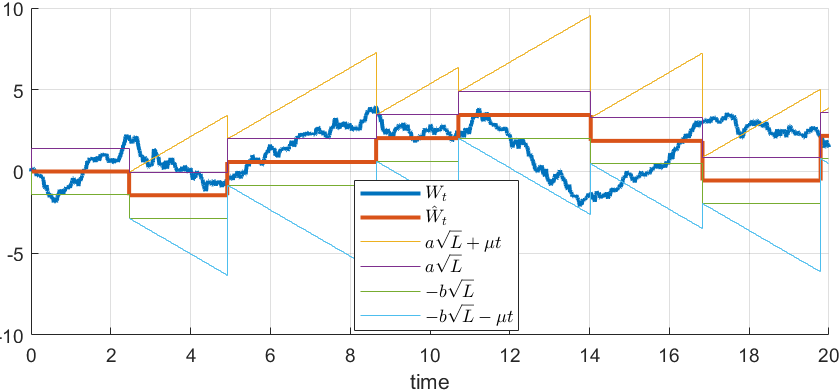}
    \caption{A sample path with $a=b=\mu=1$ and $L_n=L=2$.}
    \label{fig:sample_path}
\end{figure}

In all four events, we will be transmitting the codeword associated with the event instead of encoding and transmitting the actual value of the Wiener process. Therefore, in general, we will be taking the $(n+1)$th sample after $D_n$, at the stopping time defined $\tau(X_n)=\min\{\tilde{\tau}(X_n),\hat{\tau}_a(X_n),\hat{\tau}_b(X_n)\}$ and encode and transmit the event that resulted in the stoppage. A sample path of how $W_t$ and $\hat{W}_t$ change has been illustrated with Fig.~\ref{fig:sample_path}. Further, we will see in the next section, that the probabilities of these four events are identical and independent of previous sampling instances (see Lemma \ref{lem:iid}) and hence the same codebook can be used for encoding all the samples.  

\begin{remark}
    When $\mu=0$ (i.e., if we have only the constant thresholds instead of monotone function thresholds), if the process $X_n>a\sqrt{L_n}$ or $X_n<-b\sqrt{L_n}$, then the expected hitting time of the increment process with one of the constant thresholds will be infinite. Thus, the use of monotone function thresholds is necessary for a stable sampling procedure. Moreover, $\mu t$ can be replaced with any increasing function $f(t)$, which leads to a well-behaved stopping time, e.g., $f(t)=\mu t^2$ or $f(t)=e^{\mu t}$, etc. 
\end{remark}

\section{Main Results}
In this section, we state the main analytical results when implementing the monotone function thresholding scheme. The very first result states that under the described scheme, since we appropriately scale the thresholds based on our previous source coding length, the resulting generation of source coding lengths will be i.i.d. The proof of Lemma \ref{lem:iid} is given in Appendix \ref{appen:iid}.

\begin{lemma}\label{lem:iid}
    The source coding lengths are generated independently and identically. 
\end{lemma}

Next, using Lemma \ref{lem:iid} and the optional stopping theory, we can find $\e[\Delta_{mse}]$ as stated in Theorem \ref{thrm:mse_main}. 

\begin{theorem}\label{thrm:mse_main}
     The MSE of the Wiener process under the monotone function thresholding scheme is given by,
    \begin{align}\label{eqn:mse_main}
        \e[\Delta_{mse}]=\frac{\e[Y^4]+6\e[C(Y)Y^2]}{6\e[\tau(X_L)+L]},
    \end{align}
    where $Y=X_L+B(\tau(X_L))$, $B(t)$ is a standard Brownian motion rooted at zero, $C(Y)$ is the source coding associated with $Y$, $X_L\sim\mathcal{N}(0,L)$ and $L$ is the random variable associated with code lengths.
\end{theorem}

The proof of Theorem \ref{thrm:mse_main} can be found in Appendix \ref{appen:mse_thrm}. Next, let $p_i$ be the probability that source coding length $l_i$ is generated for the next sample (see Appendix \ref{appen:iid} for a formal description of $p_i$s). Then, for a given set of thresholds and for large $\mu$, Theorem \ref{thrm:mse_main_2} gives the closed-form expression for $\e[\Delta_{mse}]$. For the rest of the paper, we will assume $\mu$ is sufficiently large.

\begin{theorem}\label{thrm:mse_main_2}
    For sufficiently large $\mu$, $\e[\Delta_{mse}]$ is given by,
    \begin{align}\label{eqn:mse_simp}
        \e[\Delta_{mse}]=K\frac{\e_P[L^2]}{\e_P[L]}+\e_{\tilde{P}}[L],
    \end{align}
    where
    \begin{align}
        K&=\frac{3+p_2a^4+p_3b^4-\tilde{X}}{6(p_2a^2+p_3b^2+\tilde{A}+\tilde{B})},\\
        \tilde{A}&=\frac{1}{\sqrt{2\pi}}\int_{a}^{\infty}x^2e^{-\frac{x^2}{2}}\,dx,\\
        \tilde{B}&=\frac{1}{\sqrt{2\pi}}\int_{b}^{\infty}x^2e^{-\frac{x^2}{2}}\,dx,\\
        \tilde{X}&=\frac{1}{\sqrt{2\pi}}\int_{-b}^{a}x^4e^{-\frac{x^2}{2}}\,dx,
    \end{align}
 and $P=\{p_1,p_2,p_3,p_4\}$ and $\tilde{P}=\{\frac{\tilde{A}}{D},\frac{p_2a^2}{D},\frac{p_3b^2}{D},\frac{\tilde{B}}{D}\}$ with $D=p_2a^2+p_3b^2+\tilde{A}+\tilde{B}$, are PMFs on source coding lengths $\{l_1,l_2,l_3,l_4\}$. Moreover, the sampling rate (SR) of the scheme is given by,
 \begin{align}
     \text{SR}=\frac{1}{\e[\tau(X_L)+L]}=D\e_P[L].
 \end{align}
\end{theorem}

\begin{remark}
Note that the expression given in \eqref{eqn:mse_simp} is analogous to the average of age of a process whose inter arrival times between updates are governed by the random variable $L$ similar to the work in \cite{optimal_codes}. 
\end{remark}

\begin{remark}
    When $\mu$ is sufficiently large and $l_i=2$ for all $i$, our scheme effectively reduces to the optimal sampling scheme of Wiener process given in \cite{Yin_Sun} under a constant transmission delay of $2$.
\end{remark}

\begin{remark}
    Note that $\mu$ can be as large as desired, however it cannot be infinite, since this will render the process of estimating the actual value of the Wiener process using the time difference of the samples useless. No information about the actual value of the increments can be transmitted if we set $\mu=\infty$.
\end{remark}

\begin{remark}
    From \eqref{eqn:mse_simp}, it is evident that $\e[\Delta_{mse}]$ scales linearly with the bit transmission delay.
\end{remark}

\begin{remark}
    If the Wiener process has a variance of $\sigma^2$, replacing $a$, $b$ and $\mu$ with $\sigma a$, $\sigma b$ and $\sigma \mu$, respectively, linearly scales $\e[\Delta_{mse}]$ with respect to $\sigma^2$.
\end{remark}

\section{Optimal Source Codes} \label{sec:opt}
In this section, we will find the optimal source coding lengths to minimize $\e[\Delta_{mse}]$ subject to a sampling rate constraint. Now, the task at hand is two folds. First, for given values of $a$ and $b$, we need to find a set of prefix codeword lengths that minimizes \eqref{eqn:mse_simp}. This is equivalent to minimizing \eqref{eqn:mse_simp} with respect to $l_i$ subject to the Kraft's inequality. Next, we need to find the optimal values for the thresholds. Thus, our problem can be posed as the following optimization problem,
\begin{mini}
    {a,b>0, \ l_i\in \mathbbm{N}}{\frac{K\e_P[L^2]}{\e_P[L]}+\e_{\tilde{P}}[L]}
    {\label{eqn:opt_1}}
    {}
    \addConstraint{\sum_{i=1}^4 2^{-l_i}}{\leq 1}
    \addConstraint{\e_P[L]}{\geq \frac{1}{Df_{max}}},
\end{mini}
where $f_{max}$ is the maximum sampling rate allowed.

Unlike in traditional source coding problems where the objective is a linear function of coding lengths, in here, the objective function is quadratic with respect to source coding lengths. Now, instead of solving the optimization problem \eqref{eqn:opt_1} exactly, we will  solve for a relaxed version of the optimization problem by considering $l_i\in \mathbb{R}$ instead of restricting to non-negative integers. When assuming $l_i\in\mathbb{R}$, there is no real merit in considering two distinct values for $a$ and $b$ due to the symmetry of the problem. Therefore, we will consider $a=b$ for the rest of problem. Further, $a=b$ implies that $l_1=l_4$, $l_2=l_3$, $p_1=p_4$ and $p_2=p_3$. For a fixed $a$, we will first find the best source coding lengths and later optimize with respect to $a$. Thus, the problem at hand, reduces to the following,
\begin{mini}
    {l_1,l_2\in \mathbbm{R}}{\frac{K\e_P[L^2]}{\e_P[L]}+\e_{\tilde{P}}[L]}
    {\label{eqn:opt_relax}}
    {}
    \addConstraint{\sum_{i=1}^2 2^{-l_i}}{\leq \frac{1}{2}}
    \addConstraint{\e_P[L]}{\geq \frac{1}{Df_{max}}}.
\end{mini}

Then, using the Dinkelbach method \cite{dinkelbach}, we will linearize the fractional objective function as follows,
\begin{mini}
    {l_i\in \mathbbm{R}}{K\e_P[L^2]+\e_P[L]\e_{\tilde{P}}[L]-\theta\e_P[L]}
    {\label{eqn:opt_lin}}
    {}
\end{mini}
where $\theta \in \mathbb{R}$. This is optimized with respect to the same constraints as in \eqref{eqn:opt_relax}. Let $J(\theta)$ denote the optimal value of the above optimization and $J$ denote the optimal value of  \eqref{eqn:opt_relax}. Then, it is well-known that when $J(\theta)\lesseqgtr0 \iff \theta \lesseqgtr J$ \cite{Yin_Sun}.  The above optimization problem satisfies the linear independence constraint qualification (LIQC) and hence the Karush-Khun-Tucker (KKT) conditions give the necessary conditions for optimality. Let the Lagrangian of the above optimization problem be defined as,
\begin{align}
    \mathcal{L}(l,\lambda)=&K\e_P[L^2]+\e_P[L]\e_{\tilde{P}}[L]-\theta\e_P[L]\nonumber\\
    &+\gamma\bigg(\frac{1}{Df_{max}}-\e_P[L]\bigg)+\lambda\bigg(\sum_{i=1}^{2}2^{-l_i}-\frac{1}{2}\bigg),
\end{align}
where $\gamma,\lambda\geq0$ are the Lagrangian multipliers. Then, the KKT conditions are given by,
\begin{align}
    \pdv{L}{l_i} = &4Kl_ip_i+2\tilde{p}_i\e_P[L]+2p_i\e_{\tilde{P}}[L]-2(\theta+\gamma) p_i \nonumber\\
    &-\ln{2}\lambda 2^{-l_i} = 0,\label{eqn:diff}
\end{align}
with the complementary slackness conditions given by,
\begin{align}
    \lambda\bigg(\sum_{i=1}^{2}2^{-l_i}-\frac{1}{2}\bigg)&=0,\label{eqn:slack}\\
    \gamma\bigg(\frac{1}{Df_{max}}-\e_P[L]\bigg)&=0.\label{eqn:slack_2}
\end{align}

In here, $\tilde{p}_i$ is simply the $i$th element of the PMF $\tilde{P}$ defined in Theorem \ref{thrm:mse_main_2}. Next, Theorem \ref{thrm:Kraft} states an important characteristic of the optimal source coding lengths.

\begin{theorem}\label{thrm:Kraft}
    For a given threshold, the optimal source coding lengths must satisfy either the Kraft inequality or the sampling rate constraint with equality. 
\end{theorem}

\begin{remark}
As a consequence of Theorem \ref{thrm:Kraft}, if $f_{max}=\infty$, then the optimal source coding lengths must satisfy the Kraft inequality with equality.
\end{remark}

The proof of Theorem \ref{thrm:Kraft} is given in Appendix \ref{appen:Kraft}. Now, to solve the $J(\theta)$ optimization, the objective function of \eqref{eqn:opt_lin} can be expressed as a quadratic program as follows,
\begin{mini}
    {\bm{l}\in \mathbbm{R}^2}{\bm{l}^T\bm{Q}\bm{l}-\bm{q}_{\theta}^T\bm{l}}
    {\label{eqn:opt_quad}}
    {}
\end{mini}
where $\bm{l}=[l_1,l_2]^T$, $\bm{q}_{\theta}=[2\theta p_1,2\theta p_2]^T$ and $\bm{Q}$ is defined as,
\begin{align}
    \bm{Q} = 2
    \begin{bmatrix}
    Kp_1+2p_1\tilde{p}_1&p_1\tilde{p}_2+p_2\tilde{p_1}\\
    p_2\tilde{p_1}+p_1\tilde{p}_2& K+2\tilde{p_2}p_2
    \end{bmatrix}.
\end{align}
The quadratic program must be solved subject to the same constraints in \eqref{eqn:opt_relax}. Note that $\bm{Q}$ is independent of $\theta$ and only dependent on $a$. A simple numerical verification shows that $\bm{Q}$ is positive semi-definite for all $a>0$. Since the constraints in the optimization \eqref{eqn:opt_relax} are also convex, the above quadratic program can be easily solved. This gives us $J(\theta)$. Now, the optimal $\theta$ which is the solution for $J(\theta)=0$, can be found using a simple bisection search. Then, finally, we perform a one-dimensional exhaustive search to find the optimal $a$.   

\section{Numerical Results}
In this section, we validate our analysis via simulations and compare our proposed scheme with several benchmark schemes. For simulations, we discretize time $t$ into time intervals of length $\varepsilon=10^{-2}$, and we construct the Wiener process at these discretized time indices as $W_{t_i}=W_{t_{i-1}}+r$, where $r\sim\mathcal{N}(0,\varepsilon)$. In each time index, if there is no ongoing transmission, we check if the Wiener process has reached one of the thresholds described in Section~\ref{sec:mono}. In the event that it crosses one of the given thresholds, the event is source coded with $L$ bits and transmitted over $\text{round}(L/\varepsilon)$ time indices. In here, the round operation ensures that the transmission duration takes an integer multiple of our discrete time interval $\varepsilon$. After transmission, the transmitted event is decoded and the estimate $\hat{W}_t$ is updated. This process is repeated over the time horizon of $T=10^5$ (i.e., $\text{round}(T/\varepsilon)$ time indices). 

\begin{figure}[t]
    \begin{center}
    \subfigure[]{\includegraphics[width=0.475\columnwidth]{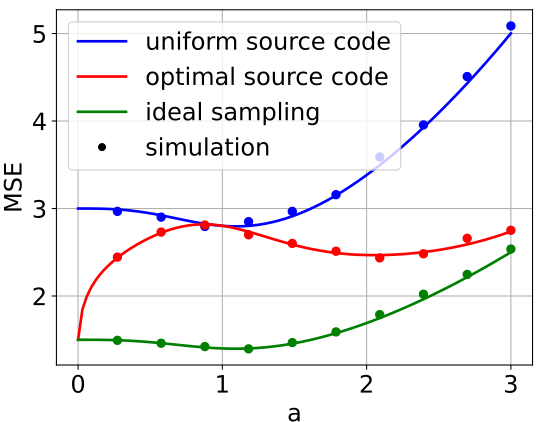}} ~ 
    \subfigure[]{\includegraphics[width=0.475\columnwidth]{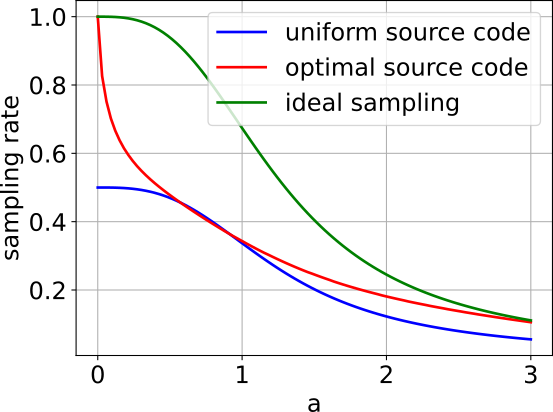}}  
    \end{center}  
    \caption{Comparison of optimum source code scheme under no sample rate constraint with benchmark schemes. Curves are obtained from analytical expressions, and dots represent simulation results.}
    \label{fig:sim_1}
\end{figure}

Fig.~\ref{fig:sim_1} illustrates the comparison of our proposed scheme with the benchmark schemes when there is no sampling rate constraint. In line with our symmetry argument in Section~\ref{sec:opt}, we fix $a=b$, $\ell_1=\ell_4$, and $\ell_2=\ell_3$. Then, we obtain the optimal $\ell_1$ and $\ell_2$ that minimize the MSE for each threshold $a$. The \emph{uniform source code} is a special case of our monotone thresholding scheme, where each event is encoded with 2 bits (i.e., $\ell_i=2$, $\forall i$). Therefore, in this case, each transmission takes 2 time units. In the \emph{ideal sampling} scheme we assume that the source is able to transmit the real-value without any source coding in unit time. In here, the source samples and transmits whenever the Wiener process exceeds the threshold $a$ when the channel is free. As suggested in \cite{Yin_Sun}, even in the absence of a sampling rate constraint, zero-wait sampling (i.e., $a=0$) is not optimal for the described benchmark schemes. Therefore, as illustrated, the minimum  MSE is achieved with a non-zero $a$ for these benchmark schemes. On the other hand, in the absence of a sampling rate constraint, $a=0$ is optimum for the optimal source coding scheme. 

Fig.~\ref{fig:sim_1}(a), compares MSE values of the three schemes where the solid curves represent the analytical results, and dots correspond to simulation results. First, we observe that the analytical curves fit with the simulation results, verifying our analysis. Additionally, we observe that the MSE performance of the optimal source coding scheme is bounded by the other two schemes. In general, when $a$ is small,  $p_1$ is also small. Therefore, the MSE is minimized by selecting $\ell_1$ close to $1$, and setting $\ell_2$ to a large value. Notice that when threshold $a$ approaches $0$, it makes $p_1=0$ and  \eqref{eqn:mse_simp} reduces to $\frac{3}{2}\ell_1$. This is minimized by setting $\ell_1=1$ and $\ell_2=\infty$ which also sets the sampling rate to  1. At this point the ideal sampling scheme and our optimal source coding scheme are identical and hence the MSEs are equivalent.

\begin{figure}[t]
    \begin{center}
    \subfigure[]{\includegraphics[width=0.475\columnwidth]{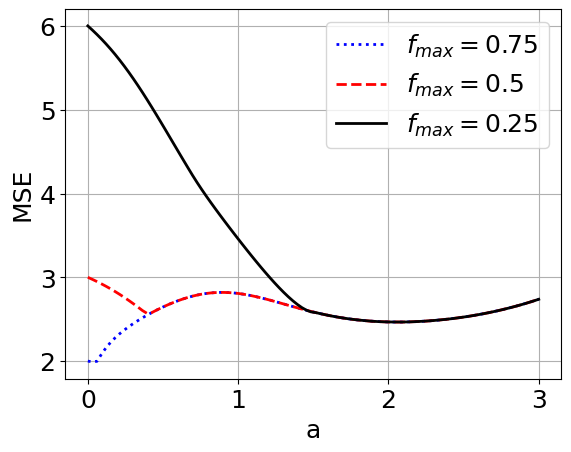}} ~ 
    \subfigure[]{\includegraphics[width=0.475\columnwidth]{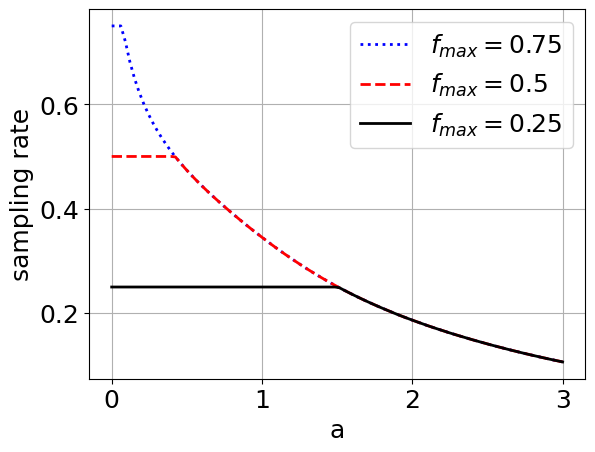}}  
    \end{center}  
    \caption{a) MSE and b) sampling rate under different sampling rate constraints.}
    \label{fig:sim_2}
\end{figure}

As Fig.~\ref{fig:sim_1} suggests, small $a$ values cause a high sampling rate for the optimal source coding in the absence of  a sampling rate constraint. Therefore, in Fig~\ref{fig:sim_2}, we illustrate the optimum MSE values under different sampling rate constraints and their corresponding sampling rates in Fig.~\ref{fig:sim_2}(a) and Fig.~\ref{fig:sim_2}(b), respectively. As highlighted in Theorem~\ref{thrm:Kraft}, optimal source coding scheme must satisfy at least one of the sampling rate and Kraft's constraints with equality. Therefore, in Fig~\ref{fig:sim_2}(b), we can identify two regions for $a$, one in which the sampling rate constraint  is active (for small $a$) and the other in which the Kraft inequality is active (for large $a$). At the intersecting point of these regions, both of these constraints are active. In addition, when $a$ is small, the MSE and the sampling rate values are mainly determined by $\ell_1$. Since the sampling rate constraint is active in this region, $\ell_1$ cannot be chosen arbitrarily close to $1$ anymore, and hence the optimal $a$ value is non-zero now.

\section{Conclusion}
We introduced a novel sampling and encoding strategy for monitoring a Wiener process. For this scheme, we have found the optimal source code lengths which suggests that a zero-wait sampling policy is optimal in the absence of a sampling rate constraint. In the presence of a sampling rate constraint, a non-zero threshold scheme with monotone function thresholds is deemed to be optimal. The effect of multiple thresholds along with monotone function thresholds is an interesting research direction which is left for future work. 

\appendices

\section{Proof of Lemma \ref{lem:iid}}\label{appen:iid}
Let $L_n$ be the source coding length of the previous sample. Let $X_n=W_{D_n}-W_{S_n}$. If $X_n>a\sqrt{L_n}$, then $L_{n+1}=l_1$ and if $X_n<-b\sqrt{L_n}$ then $L_{n+1}=l_4$. If $X_n=x$ for $x\in(-b\sqrt{L_n},a\sqrt{L_n})$, then the $W_t-S_{n}$ will hit the upper constant threshold with probability, $\frac{(x+b\sqrt{L_n})}{\sqrt{L_n}(a+b)}$ and would hit probability $\frac{(a\sqrt{L_n})-x}{\sqrt{L_n}(a+b)}$. If it hits the upper threshold we will have $L_{n+1}=l_2$ and if hits the lower threshold we will have $L_{n+1}=l_3$. Also note that $X_n\sim \mathcal{N}(0,L_n)$. Therefore, $L_{n+1}$ will take the values in $\{l_1,l_2,l_3,l_4$\} with the following probabilities,
\begin{align}
    \mathbb{P}(L_{n+1}=l_1|L_{n})&=\frac{1}{\sqrt{2\pi L_n}}\int_{a\sqrt{L_n}}^{\infty}e^{-\frac{x^2}{2L_n}}\,dx\nonumber\\
    &=\frac{1}{\sqrt{2\pi}}\int_{a}^{\infty}e^{-\frac{x^2}{2}}\,dx\nonumber\\
    &=p_1,\\
    \mathbb{P}(L_{n+1}=l_2|L_{n})&=\frac{1}{\sqrt{2\pi L_n}}\int_{-b\sqrt{L_n}}^{a\sqrt{L_n}}\frac{(x+b\sqrt{L_n})}{\sqrt{L_n}(a+b)}e^{-\frac{x^2}{2L_n}}\,dx\nonumber\\
    &=\frac{1}{\sqrt{2\pi}}\int_{-b}^{a}\frac{(x+b)}{(a+b)}e^{-\frac{x^2}{2}}\,dx\nonumber\\
    &=p_2,\\
    \mathbb{P}(L_{n+1}=l_3|L_{n})&=\frac{1}{\sqrt{2\pi L_n}}\int_{-b\sqrt{L_n}}^{a\sqrt{L_n}}\frac{(a\sqrt{L_n}-x)}{\sqrt{L_n}(a+b)}e^{-\frac{x^2}{2L_n}}\,dx\nonumber\\
    &=\frac{1}{\sqrt{2\pi}}\int_{-b}^{a}\frac{(a-x)}{(a+b)}e^{-\frac{x^2}{2}}\,dx\nonumber\\
    &=p_3,\\
    \mathbb{P}(L_{n+1}=l_4|L_{n})&=\frac{1}{\sqrt{2\pi L_n}}\int_{-\infty}^{-b\sqrt{L_n}}e^{-\frac{x^2}{2L_n}}\,dx\nonumber\\
    &=\frac{1}{\sqrt{2\pi}}\int_{-\infty}^{-b}e^{-\frac{x^2}{2}}\,dx\nonumber\\
    &=p_4.
\end{align}
As seen above, the conditional probabilities of $L_{n+1}$ are independent of $L_n$, thus proving the result.

\section{Proof of Theorem \ref{thrm:mse_main}}\label{appen:mse_thrm}
Let us first state an important lemma derived from the optional stopping theory of Wiener process whose proof is available in \cite{Yin_Sun} and \cite{Serfozo}.

\begin{lemma}\label{lem:mtng}
    Let $\tau$ be a stopping time of the Wiener process with $\e[\tau]<\infty$. If either $\sup_{0\leq t\leq \tau}|W_t|<\infty$ or if $\e[\tau^2]<\infty$, then we have from optional stopping theorem and Doob's $L^p$ maximal inequality,
    \begin{align}
        \e\left[\int_{0}^{\tau}W_t^2\,dt\right] = \frac{\e[W_{\tau}^4]-\e[W_0^4]}{6}.
    \end{align}
\end{lemma}

Let $X_n=W_{D_n}-W_{S_n}$. Then, the MSE from $D_n$ to $D_{n+1}$ can be obtained as follows,
\begin{align}
    &\e\left[\int_{D_{n}}^{D_{n+1}} (W_t-\hat{W}_t)^2\,dt\right]\nonumber\\
    &=\e\left[\int_{D_{n}}^{D_n+\tau(X_n)+L_{n+1}} (W_t-W_{D_n}+W_{D_n}-W_{S_n})^2\,dt\right]\\
    &=\e\left[\int_{0}^{\tau(X_n)+L_{n+1}} (W_{t+D_n}-W_{D_n}+W_{D_n}-W_{S_n})^2\,dt\right]\\
    &=\e\left[\int_{0}^{\tau(X_n)+L_{n+1}} (B(t)+X_n)^2\,dt\right].
\end{align}
Since $X_n\sim\mathcal{N}(0,L_n)$ and $L_n$ are i.i.d.,  from renewal reward theorem (RRT), we have the following,
\begin{align}\label{eqn:mse_gen}
    \e[\Delta_{mse}]=\frac{\e\left[\int_{0}^{\tau(X)+\tilde{L}} (B(t)+X)^2\,dt\right]}{\e[\tau(X)+L]},
\end{align}
where $X\sim \mathcal{N}(0,L)$, $L$ is the discrete random variable associated with the source code lengths of the previous sample and $\tilde{L}$ is the source coding length of the next sample which has the same distribution as $L$.
    
Now, let $\tau(X)=\tau_{x}$ and $Y_t=B(t)+X$. Then, from Lemma \ref{lem:mtng}, we have,
\begin{align}
     6\e&\left[\int_{0}^{\tau(X)+\tilde{L}} (B(t)+X)^2\,dt\right]\nonumber\\        &=6\e\left[\int_{0}^{\tau_{x}+\tilde{L}}Y_t^2\,dt\right]\\
     &=\e[Y_{\tau_{x}+\tilde{L}}^4]-\e[X^4]\\
     &=\e[(Y_{\tau_{x}+\tilde{L}}-Y_{\tau_{x}}+Y_{\tau_{x}})^4]-3\e[L^2]\\
     &=\e[(Y_{\tau_{x}+\tilde{L}}-Y_{\tau_{x}})^4]+6\e\left[(Y_{\tau_{x}+\tilde{L}}-Y_{\tau_{x}})^2Y_{\tau_{x}}^2\right]\nonumber\\
     &\quad+\e[Y_{\tau_{x}}^4]-3\e[L^2]\\
     &=3\e[\tilde{L}^2]+6\e[\tilde{L}Y_{\tau_{x}}^2]+\e[Y_{\tau_{x}}^4]-3\e[L^2]\\
     &=\e[Y_{\tau_{x}}^4]+6\e[C(Y_{\tau_{x}})Y_{\tau_{x}}^2].
\end{align}
 
\section{Proof of Theorem \ref{thrm:mse_main_2}}
Before proceeding with the computation of $\e[\Delta_{mse}]$, we will first state two useful properties of the considered stopping times, which will be  helpful towards proving the main result. The proof of Lemma \ref{lem:mnts} can be found in \cite{Serfozo}.

\begin{lemma}\label{lem:mnts}
    For $\mu>0$, let $\tau_c$ be the hitting time of a Brownian process  $\mu t+B(t)$ rooted at zero, with level $c>0$. Then the Laplace transform $\Psi(\lambda)$ of the distribution of $\tau_c$ and the moments of $\tau_c$ will be given by,
    \begin{align}
         \Psi(\lambda)=\e[e^{-\lambda \tau_c}]=e^{-c(\sqrt{\mu^2+2\lambda}-\mu)},\label{eqn:lap_tau}\\
          \e[\tau_c^k]=(-1)^k\pdv[k]{\Psi}{\lambda}\Bigg|_{\lambda=0}.
    \end{align}
\end{lemma}

\begin{corollary}\label{cor:mnts}
     For $\mu>0$, let $\tau_c$ be the hitting time of a Brownian process  $\mu t+B(t)$ rooted at zero, with level $c>0$. Then, the first 4 moments will be given by,
    \begin{align}
         \e[\tau_c]&=\frac{c}{\mu},\\
         \e[\tau_c^2]&=\frac{c^2}{\mu^2}+\frac{c}{\mu^3},\\
         \e[\tau_c^3]&=\frac{c^3}{\mu^3}+\frac{3c^2}{\mu^4}+\frac{3c}{\mu^5},\\
         \e[\tau_c^4]&=\frac{c^4}{\mu^4}+\frac{6c^3}{\mu^5}+\frac{15c^2}{\mu^6}+\frac{15c}{\mu^7}.
    \end{align}
\end{corollary}

Next, define by $A_k$ and $B_k$ the following integrals,
\begin{align}
    A_k&=\frac{1}{\sqrt{2\pi}}\int_{a}^{\infty}\left(x-a\right)^ke^{-\frac{x^2}{2}}\,dx,\\
    B_k&=\frac{1}{\sqrt{2\pi}}\int_{b}^{\infty}\left(x-b\right)^ke^{-\frac{x^2}{2}}\,dx.
\end{align}

Now, the MSE can be obtained as follows,
\begin{align}
    \e[C(Y)&Y^2|X>a\sqrt{L},X,L]\nonumber\\
    &=l_1\e[(a\sqrt{L}+\mu\tau_{x})^2]\\
    &=l_1(a^2L+2a\mu\sqrt{L}\e[\tau_{x}]+\mu^2\e[\tau_{x}^2]),
\end{align}
where $\tau(X)=\tau_x$ is used for brevity. Now, note that $\tau_x$ for $x>a\sqrt{L}$ is equal in distribution to the hitting time of a Brownian motion with a drift of $\mu$ with level $x-a\sqrt{L}$. Therefore, we further have that,
\begin{align}
    \e[C(Y)&Y^2|X>a\sqrt{L},X,L] \nonumber\\
    =&l_1\bigg(a^2L+(2a\sqrt{L})(x_L-a\sqrt{L})\nonumber\\
    &+(x_L-a\sqrt{L})^2+\frac{(x_L-a\sqrt{L})}{\mu}\bigg),
\end{align}
where $x_L$ is a realization of $X$ given $L$. Now, accounting for the distribution of $X$ yields,
\begin{align}
    \e[&\mathbbm{1}\{X>a\sqrt{L}\}C(Y)Y^2|L]\nonumber\\
    &=l_1\bigg(p_1a^2L+(2a\sqrt{L})\sqrt{L}A_1+LA_2+\frac{\sqrt{L}A_1}{\mu}\bigg)\\
    &=\underset{\tilde{A}}{\underbrace{\big(p_1a^2+2aA_1+A_2\big)}}l_1L+l_1\frac{A_1\sqrt{L}}{\mu}\\
    &=l_1\tilde{A}L+l_1\frac{A_1\sqrt{L}}{\mu}.
\end{align}
Similarly, for $X<-b\sqrt{L}$, we have,
\begin{align}
     \e[&\mathbbm{1}\{X<-b\sqrt{L}\}C(Y)Y^2|L]\nonumber\\
     &=\underset{\tilde{B}}{\underbrace{\big(p_1b^2+2bB_1+B_2\big)}}l_4L+l_4\frac{B_1\sqrt{L}}{\mu}\\
     &=l_4\tilde{B}L+l_4\frac{B_1\sqrt{L}}{\mu}.
\end{align}
Now, considering the case $X\in (-b\sqrt{L},a\sqrt{L})$, we have,
\begin{align}
    \e[&\mathbbm{1}\{-b\sqrt{L}\leq X \leq a\sqrt{L}\}C(Y)Y^2|L] \nonumber\\
    &=L\big(p_2l_2a^2+p_3l_3b^2\big).
\end{align}
Therefore, we have,
\begin{align}
    \e[C(Y)Y^2]
    &=\e\bigg[l_1\tilde{A}L+l_1\frac{A_1\sqrt{L}}{\mu}+ l_4\tilde{B}L+l_4\frac{B_1\sqrt{L}}{\mu}\nonumber\\
    &\qquad+(p_2l_2a^2+p_3l_3b^2)L\bigg]\\
    &=(l_1\tilde{A}+l_4\tilde{B}+p_2l_2a^2+p_3l_3b^2)\e_P[L]\nonumber\\
    &\qquad+\frac{(l_1A_1+l_4B_1)}{\mu}\e_P[\sqrt{L}].
\end{align}
Similarly,  $\e[Y^4]$ can also be evaluated as follows,
\begin{align}
    \e[&Y^4|X>a\sqrt{L}, X, L]\nonumber\\
    &=(a\sqrt{L}+\mu\tau_x)^4\\
    &=a^4L^2+4a^3L\sqrt{L}\mu\e[\tau_x]+6a^2L\mu^2\e[\tau_x^2]\nonumber\\
    &\quad+4a\sqrt{L}\mu^3\e[\tau_x^3]+\mu^4\e[\tau_x^4].
\end{align}
 Now, using Corollary \ref{cor:mnts} and accounting for the distribution of $X$ yields,
\begin{align}
    \e[&\mathbbm{1}\{X>a\sqrt{L}\}Y^4|L]\nonumber\\
    =&p_1a^4L^2+4a^3L\sqrt{L}(A_1\sqrt{L})+6a^2L\bigg(A_2L+\frac{A_1\sqrt{L}}{\mu}\bigg)\nonumber\\
    &+4a\sqrt{L}\bigg(A_3L\sqrt{L}+\frac{3A_2L}{\mu}+\frac{3A_1\sqrt{L}}{\mu^2}\bigg)\nonumber\\
    &+A_4L^2+\frac{6A_3L\sqrt{L}}{\mu}+\frac{15A_2L}{\mu^2}+\frac{15A_1\sqrt{L}}{\mu^3}.
\end{align}
Similarly, we have,
\begin{align}
    \e[&\mathbbm{1}\{X<-b\sqrt{L}\}Y^4|L]\nonumber\\
    =&p_4b^4L^2+4b^3L\sqrt{L}(B_1\sqrt{L})+6b^2L\bigg(B_2L+\frac{B_1\sqrt{L}}{\mu}\bigg)\nonumber\\
    &+4b\sqrt{L}\bigg(B_3L\sqrt{L}+\frac{3B_2L}{\mu}+\frac{3B_1\sqrt{L}}{\mu^2}\bigg)\nonumber\\
    &+B_4L^2+\frac{6B_3L\sqrt{L}}{\mu}+\frac{15B_2L}{\mu^2}+\frac{15B_1\sqrt{L}}{\mu^3}.
\end{align}
Also, we have,
\begin{align}
    \e[\mathbbm{1}\{-b\sqrt{L}\leq X \leq a\sqrt{L}\}Y^4|L]=L^2(p_2a^4+p_3b^4).
\end{align}
Hence, $\e[Y^4]$ can be computed by summing the above three equations and taking the expectation with respect to $L$. 

Now, to evaluate the denominator of the \eqref{eqn:mse_gen}, we have from Corollary \ref{cor:mnts},
\begin{align}
    \e[\mathbbm{1}\{X>a\sqrt{L}\}\tau(X)]&=\frac{\sqrt{L}A_1}{\mu},\\
    \e[\mathbbm{1}\{X<-b\sqrt{L}\}\tau(X)]&=\frac{\sqrt{L}B_1}{\mu},
\end{align}
and from Wald's identity for the Wiener process, we have,
\begin{align}
    &\e\left[\mathbbm{1}\{-b\sqrt{L}\geq X\leq a\sqrt{L}\}\tau(X)\right]\nonumber\\
    &=\e\left[\e[\tau(X)|X,\mathbbm{1}\{-b\sqrt{L}\geq X\leq a\sqrt{L}\}]\right]\\
    &=\e\left[\e[Y^2-X^2|X,\mathbbm{1}\{-b\sqrt{L}\geq X\leq a\sqrt{L}\}]\right]\\
    &=(p_2a^2+p_3b^2)L-\frac{L}{\sqrt{2\pi}}\int_{-b}^ax^2e^{-\frac{x^2}{2}}\,dx.\\
\end{align}
Therefore, we have,
\begin{align}
    \e[\tau(X)+L]=&(p_2a^2+p_3b^2+(\tilde{A}+\tilde{B}))\e_P[L]\nonumber\\
    &+\frac{(A_1+B_1)\e_P[\sqrt{L}]}{\mu}.
\end{align}
Now if set $\mu$ to a large value, we have,
\begin{align}
    \e[Y^4]&=\Big((p_1a^4+4a^3A_1+6a^2A_2+4aA_3+A_4)\nonumber\\
    &\qquad\quad+(p_4b^4+4b^3B_1+6b^2B_2+4bB_3+B_4)\nonumber\\
    &\qquad\quad+(p_2a^4+p_3b^4)\Big)\e_P[L^2]\\
    &=\Big(\frac{1}{\sqrt{2\pi}}\int_a^{\infty}x^4e^{-\frac{x^2}{2}}\,dx+\frac{1}{\sqrt{2\pi}}\int_b^{\infty}x^4e^{-\frac{x^2}{2}}\,dx\nonumber\\
    &\qquad\quad+(p_2a^4+p_3b^4)\Big)\e_P[L^2]\\
    &=\Big(3+p_2a^4+p_3b^4-\frac{1}{\sqrt{2\pi}}\int_{-b}^{a}x^4e^{-\frac{x^2}{2}}\,dx\Big)\e_P[L^2]\\
    &=(3+p_2a^4+p_3b^4-\tilde{X})\e_P[L^2],
\end{align}
and 
\begin{align}
    \e[C(Y)Y^2]=(l_1\tilde{A}+l_4\tilde{B}+p_2l_2a^2+p_3l_3b^2)\e_P[L].
\end{align}
Further, $\e[\tau(X)+L]$ simplifies to the following,
\begin{align}
    \e[\tau(X)+L]&=(p_2a^2+p_1b^2+\tilde{A}+\tilde{B})\e_P[L].
\end{align}
Thus, $\e[\Delta_{mse}]$ will be given by,
\begin{align}
    \e[\Delta_{mse}]=&\frac{(3+p_2a^4+p_3b^4-\tilde{X})\e_P[L^2]}{6(p_2a^2+p_3b^2+\tilde{A}+\tilde{B})\e_P[L]}\nonumber\\
    &+\frac{l_1\tilde{A}+l_4\tilde{B}+p_2l_2a^2+p_3l_3b^2}{p_2a^2+p_3b^2+\tilde{A}+\tilde{B}}\\
    =&K\frac{\e_P[L^2]}{\e_P[L]}+\e_{\tilde{P}}[L].
\end{align}

\section{Proof of Theorem \ref{thrm:Kraft}}\label{appen:Kraft}
Assume at the optimal $\theta$, we have $\sum_{i=1}^{2}2^{-l_i}<\frac{1}{2}$ and $\e_P[L]>\frac{1}{Df_{max}}$. Then, from \eqref{eqn:slack} and \eqref{eqn:slack_2}, we have $\gamma=\lambda=0$. Then, from \eqref{eqn:diff}, we have,
\begin{align}
    2Kl_ip_i+\tilde{p}_i\e_P[L]+p_i\e_{\tilde{P}}[L]=\theta p_i. \label{eqn:diff2}
\end{align}
Then, summing both sides with respect to $i$, we have,
\begin{align}
    (2K+1)\e_P[L]+\e_{\tilde{P}}[L]=\theta.
\end{align}
Then, from \eqref{eqn:diff2}, we can obtain that,
\begin{align}
    l_i&=\left(2K+1-\frac{\tilde{p}_i}{p_i}\right)\frac{\e_P[L]}{2K}\\
    &=\left(1+\frac{p_i-\tilde{p}_i}{2Kp_i}\right)\e_P[L].
\end{align}

Now, we can obtain $J(\theta)$ as follows,
\begin{align}
    J(\theta)&=\sum_{i=1}^4p_i\left(1+\frac{p_i-\tilde{p}_i}{2Kp_i}\right)^2(\e_P[L])^2+\e_P[L](\e_{\tilde{P}}[L]-\theta)\nonumber\\
    &=\sum_{i=1}^4p_i\left(1+\frac{p_i-\tilde{p}_i}{2Kp_i}\right)^2(\e_P[L])^2-(2K+1)(\e_{P}[L])^2\nonumber\\
    &=\underset{\tilde{K}}{\underbrace{\left(\sum_{i=1}^4p_i\left(1+\frac{p_i-\tilde{p}_i}{2Kp_i}\right)^2-(2K+1)\right)}}(\e_P[L])^2.
\end{align}
A simple numerical verification shows that $\tilde{K}<0$, for all $a\geq 0$. Since at the optimal $\theta$, we have $J(\theta)=0$, this leads to a contradiction as $\e_P[L]>0$.

\bibliographystyle{unsrt}
\bibliography{refs}
\end{document}